\documentclass[lettersize,journal]{IEEEtran}
\usepackage{amsmath,amsfonts}
\usepackage{algorithmic}
\usepackage{algorithm}
\usepackage{array}
\usepackage[caption=false,font=normalsize,labelfont=sf,textfont=sf]{subfig}
\usepackage{textcomp}
\usepackage{stfloats}
\usepackage{url}
\usepackage{verbatim}
\usepackage{graphicx}
\usepackage{cite}
\usepackage[normalem]{ulem}
\usepackage{xcolor}
\begin{document}

\title{A Compact Single-Beam Optically Pumped Magnetometer Based on Absorption Detection
}

\author{Jia-You Liou, Yi-Hsuan Chen, Hsiang-An Chuang, Chih-Hsuan Huang, Zong-Hsin Liu, I-Chiao Tsai, Kai-Hsin Peng, and Yi-Hsin Chen
\thanks{This work was supported by Grants No. 114-2112-M-110-005 of the National Science and Technology Council, Taiwan, and No. NHRI-14A1-BN-CO-47-2525-1 of the Institute of Biomedical Engineering and Nanomedicine, National Health Research Institutes, Taiwan; by the Ministry of Economic Affairs of Taiwan under Grant Nos. 115-EC-17-A-22-1905, 114-EC-17-A-22-1905, and 113-EC-17-A-22-1905.}
\thanks{Jia-You Liou, Kai-Hsin Peng, and Yi-Hsin Chen are with the Department of Physics, National Sun Yat-Sen University, Kaohsiung, 80424, Taiwan. (Corresponding author: yihsin.chen@mail.nsysu.edu.tw).}
\thanks{Yi-Hsuan Chen, Chih-Hsuan Huang, Zong-Hsin Liu, and I-Chiao Tsai are with the Metal Industries Research \& Development Centre, Kaohsiung, 811, Taiwan.
}
\thanks{Hsiang-An Chuang is with the Department of Electrical Engineering, National Sun Yat-Sen University, Kaohsiung, 80424, Taiwan.
}
}

\markboth{IEEE TRANSACTIONS ON INSTRUMENTATION AND MEASUREMENT}%
{Shell \MakeLowercase{\textit{et al.}}: A Sample Article Using IEEEtran.cls for IEEE Journals}


\maketitle

\begin{abstract}
We demonstrate a compact single-beam optically pumped magnetometer (OPM) employing a pure $^{87}$Rb microfabricated vapor cell with 500~Torr buffer gas. The residual magnetic field is actively minimized to below $400$~pT through a combination of $\mu$--metal magnetic shielding and internal compensation coils. We systematically characterize the OPM response as a function of laser detuning and optical intensity, and derive the magnetic-field sensitivity from the signal slope and noise amplitude spectral density. A best sensitivity of 11~fT$/\sqrt{\rm{Hz}}$ is achieved near the resonance transition at an optimal optical intensity, using a small cell volume of only 0.027 cm$^3$. 
Furthermore, a miniaturized OPM prototype was developed using a folded flexible printed circuit board (FPCB) to integrate the vapor cell, a compact heater, and magnetic-field coils into a cubic sensing head. Two coil configurations patterned directly on the FPCB were investigated. Under optimized operating conditions, the OPM prototype achieved a sensitivity of 66~fT$/\sqrt{\rm{Hz}}$.
These results demonstrate that high-buffer-pressure single-beam OPMs can achieve excellent sensitivity, establishing a pathway toward compact and scalable arrays.
\end{abstract}


\newcommand{\FigZFR}{
\begin{figure}[t]
\centering
\includegraphics[width=1\columnwidth]{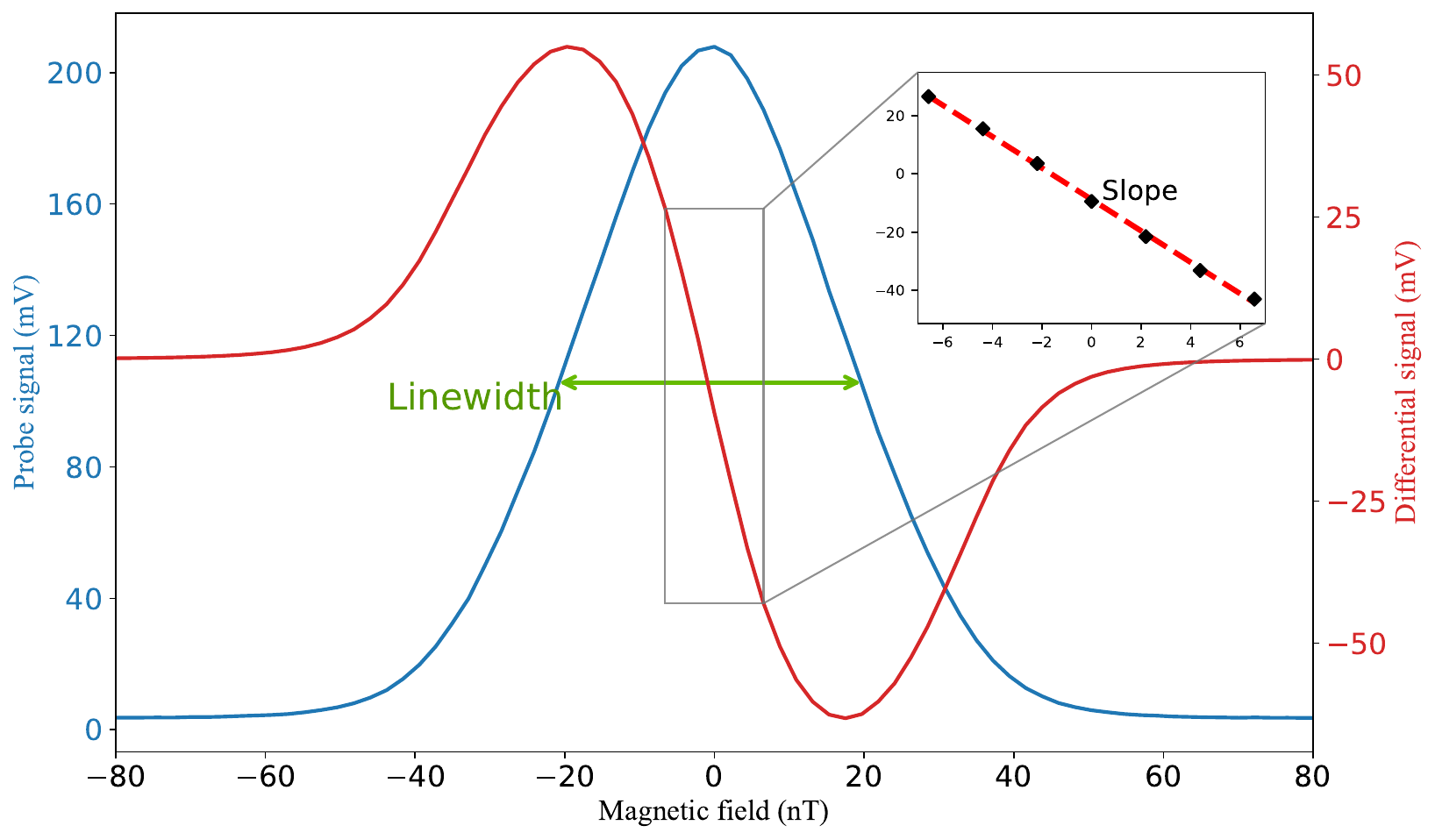}
\caption{A typical zero-field resonance curve and its differential signal as a function of the transverse magnetic field. The linear region of the differential signal near zero field is fitted to obtain the slope $dS/dB$, as shown in the inset.}
\label{fig:zfr}
\end{figure}
}

\newcommand{\Figsetup}{
\begin{figure}[t]
\centering
\includegraphics[width=1\columnwidth]{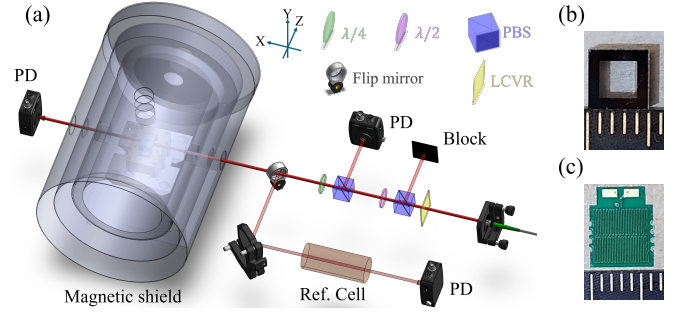}
\caption{Experimental setup. (a) The $\sigma^+$-polarized probe beam, generated by a DBR laser, is directed into a vapor cell with buffer gas 500~Torr of N$_2$ and pure $^{87}$Rb. 
The four-layer shielding passively isolates external stray magnetic fields below 400~pT. 
Within the shielding chamber, three pairs of Helmholtz coils generate the required DC bias magnetic field. 
The transmitted probe signal is detected by a photodetector (PD) and sent to a lock-in amplifier (LIA). (b) Photograph of the microfabricated vapor cell made by Twinleaf. (c) Photograph of the custom ceramic PCB heater. The double-sided counter-running circuit design reduces the generated magnetic field, while a 351~kHz AC current maintains the vapor cell at 437~K. } 
\label{fig:setup}
\end{figure}
}

\newcommand{\Figspectrum}{
\begin{figure}[t]
\centering
\includegraphics[width=1\columnwidth]{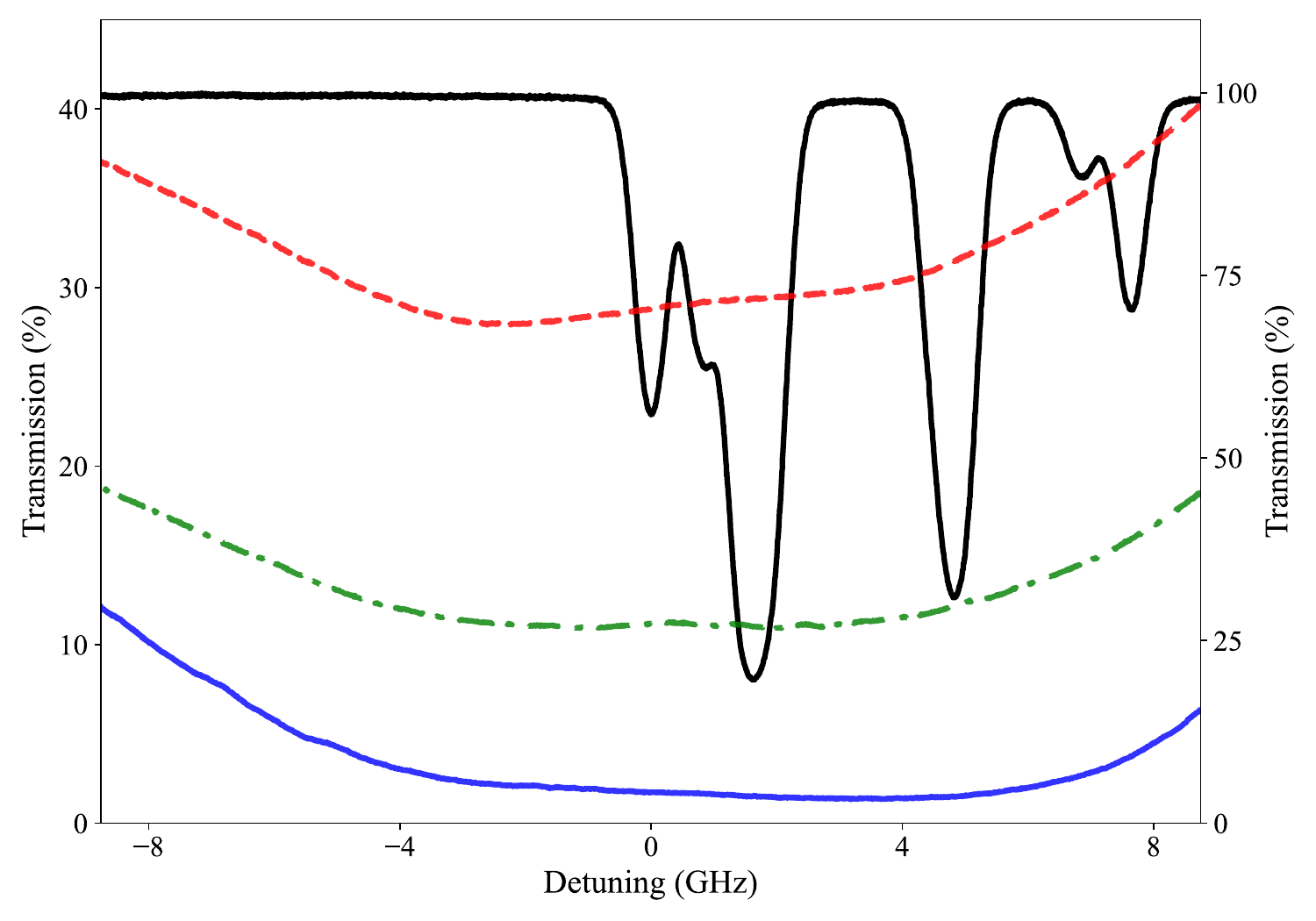}
\caption{Transmission spectra of pure $^{87}$Rb vapor in the presence of 500~Torr N$_2$ buffer gas at different optical intensities. The red (short-dashed), green (long-dashed), and blue (solid) curves correspond to intensities of 11, 6.8, and 2.9 mW/cm$^2$, respectively. The black curve shows the transmission spectrum of a reference Rb cell (right axis), used to calibrate the laser frequency.
}
\label{fig:spectrum}
\end{figure}
}

\newcommand{\Figdensity}{
\begin{figure}[t]
\centering
\includegraphics[width=1\columnwidth]{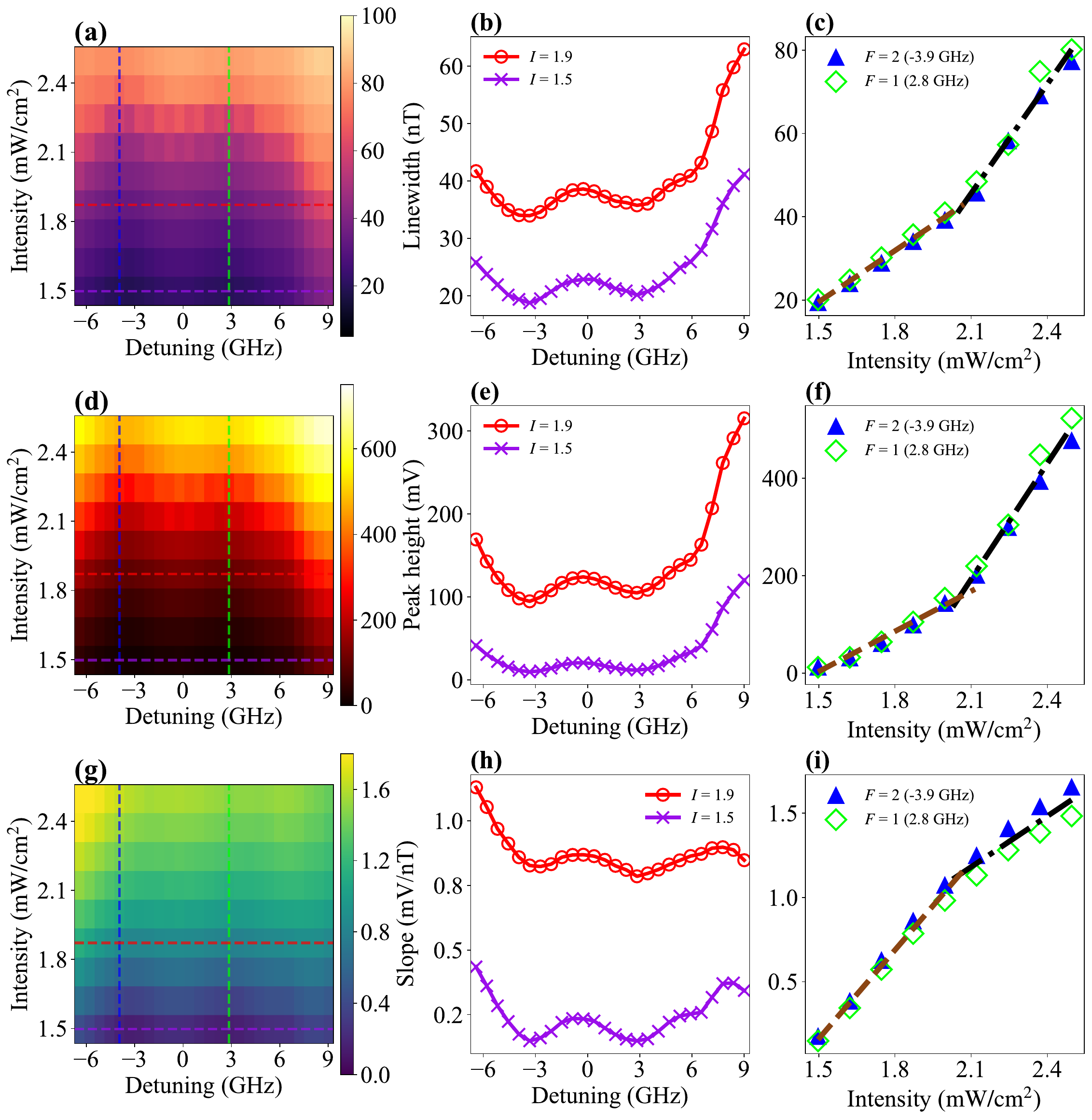}
\caption{Effects of optical intensity and laser detuning on the resonance linewidth, peak height, and differential slope. The left column shows density plots of (a) the linewidth (FWHM), (d) the peak height, and (g) the differential slope.
For each quantity, two representative optical intensities are extracted from the corresponding density plot to illustrate the dependence on laser detuning, as shown in the middle column.
Two representative detunings ($F=2$ transition at -3.9~GHz and $F=1$ transition at 2.8~GHz) are extracted to illustrate the dependence on optical intensity, as shown in the right column. The brown dashed and black dash-dotted lines are guides to the eye.
}
\label{fig:density}
\end{figure}
}

\newcommand{\Figsensitivity}{
\begin{figure}[t]
\centering
\includegraphics[width=1\columnwidth]{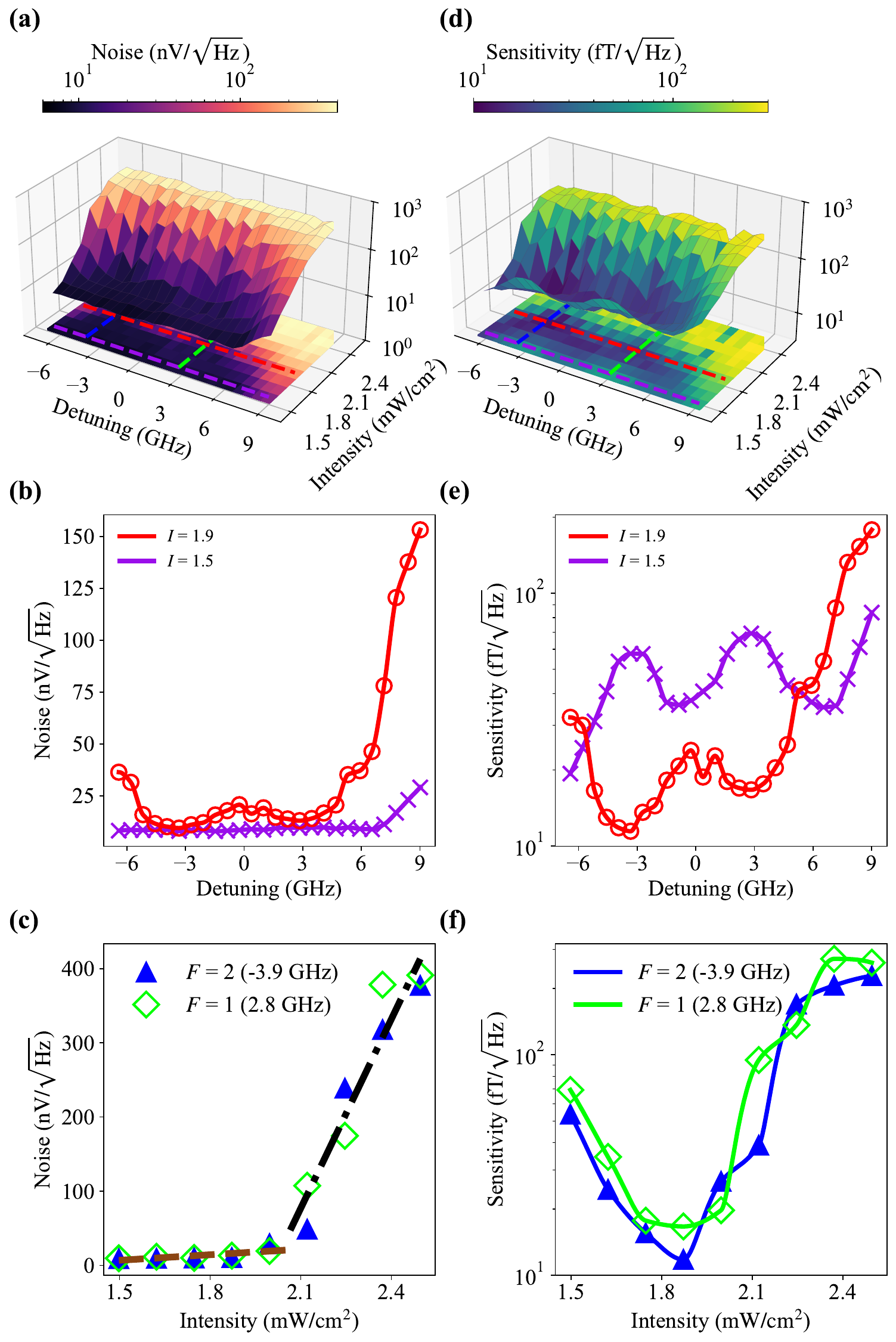}
\caption{Effects of optical intensity and laser detuning on the measured noise and OPM sensitivity.
(a)\&(d) Density plots of the noise and sensitivity as functions of optical intensity and laser detuning.
(b)\&(e) Noise and sensitivity detuning dependence at two representative optical intensities. 
(c)\&(f) Noise and sensitivity optical-intensity dependence at the two transitions.  
The best sensitivity of $11~\rm{fT}/\sqrt{\rm{Hz}}$ is obtained near the $F=2$ resonance at an optical intensity of $1.9~\rm{mW/cm^2}$. The brown dashed and black dash-dotted lines are guides to the eye.
}
\label{fig:sensitivity}
\end{figure}
}

\newcommand{\FigFPCB}{
\begin{figure}[t]
\centering
\includegraphics[width=1\columnwidth]{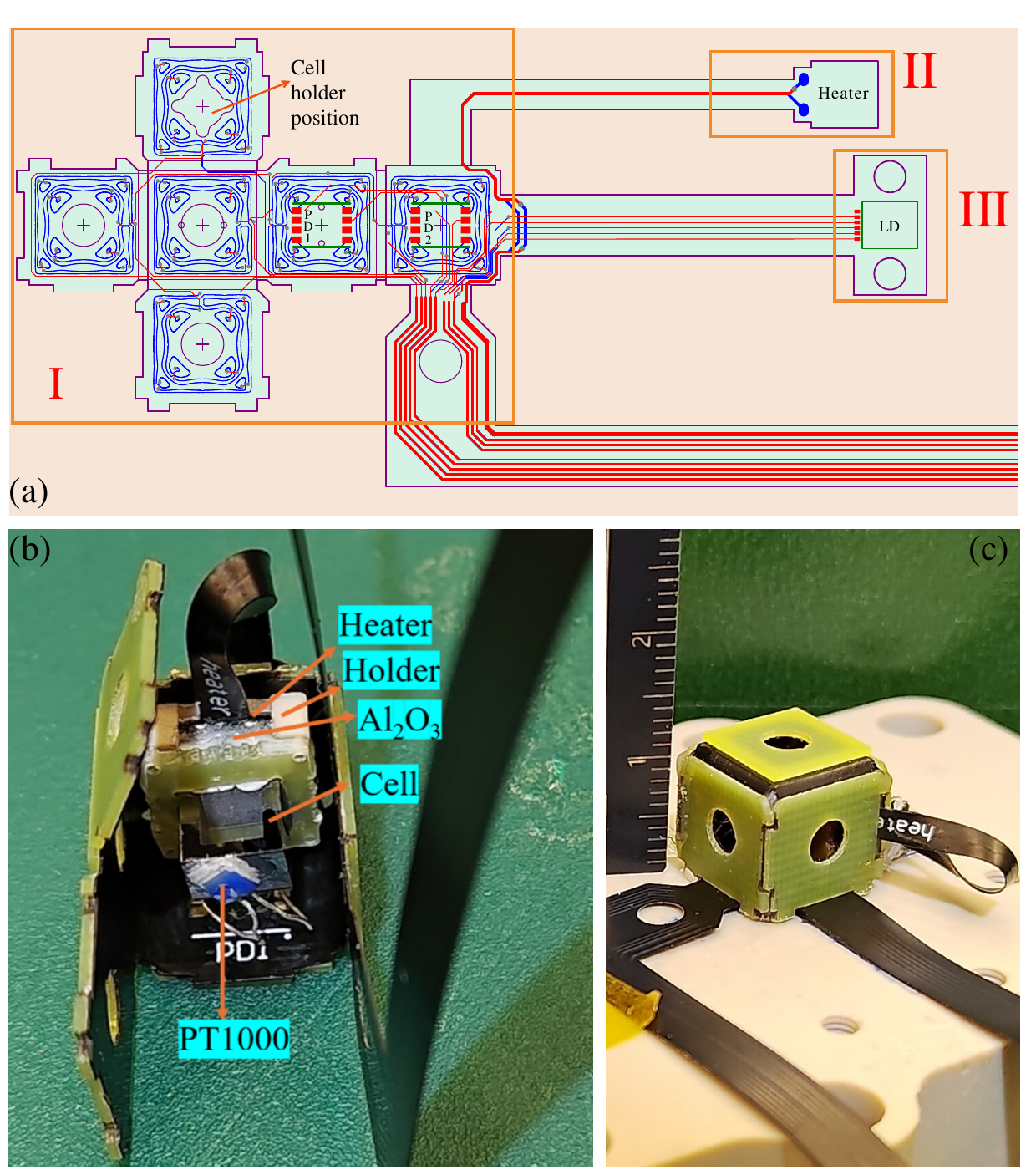}
\caption{Design and assembly of the miniaturized OPM sensor prototype.
(a) Layout of the custom integrated FPCB, including the foldable sensing region (I), heater region (II), and laser region (III), with magnetic-field coils, heater connections, and electrical traces. The PDs and laser diode will be integrated in future work.
(b) Partially assembled sensing head with the six support plates, vapor cell, heater, cell holder, Al$_2$O$_3$ plate, and PT1000 temperature sensor.
(c) Fully assembled sensing head after folding the FPCB into a compact cubic structure.}
\label{fig:FPCB}
\end{figure}
}

\newcommand{\FigBdis}{
\begin{figure}[t]
\centering
\includegraphics[width=1\columnwidth]{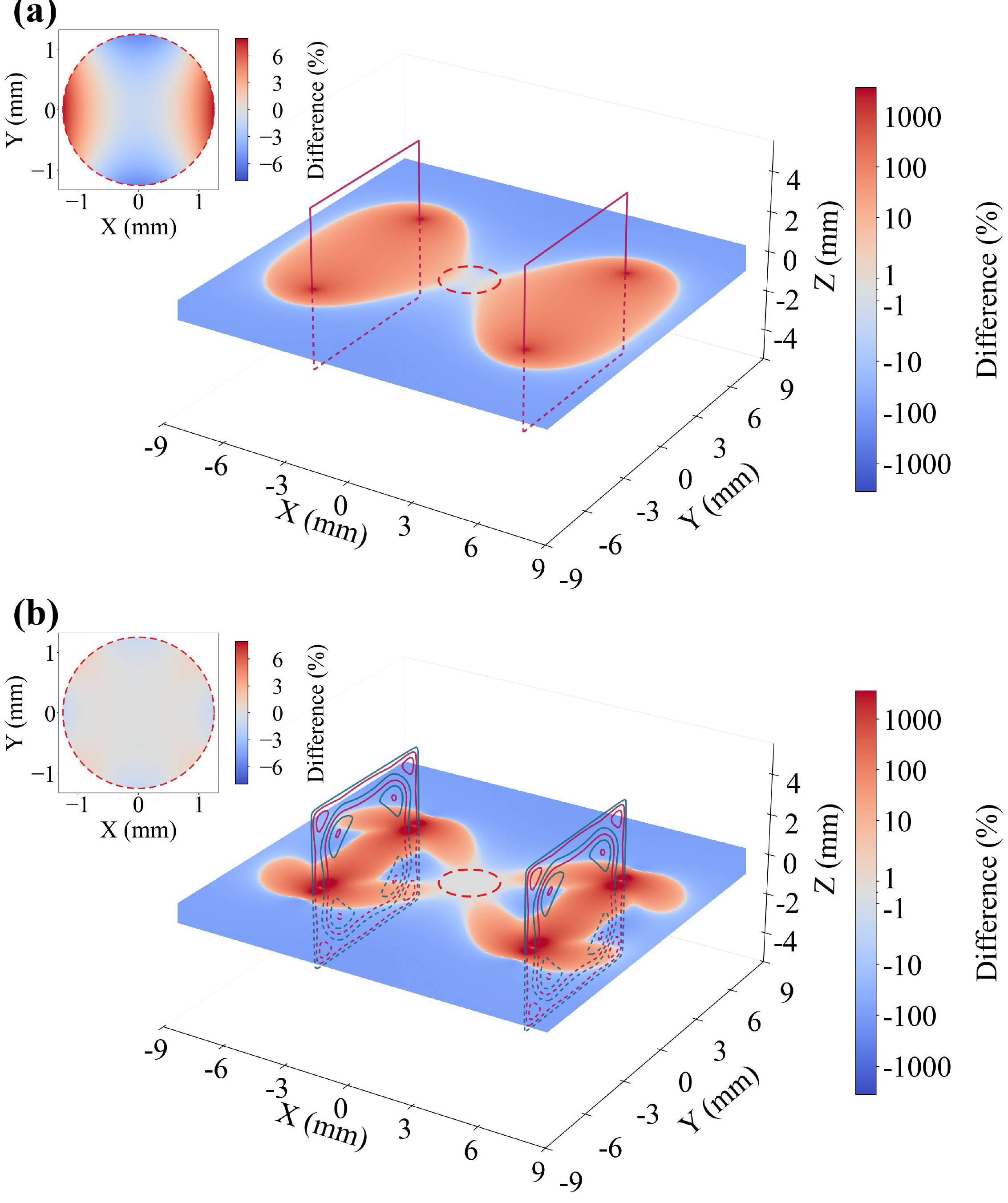}
\caption{Simulated spatial magnetic-field distributions generated by (a) SH and (b) BP configurations. The  coil current is set to $1~\rm{mA}$. In the BP design, the red and green traces indicate opposite current directions. The color scale in the density plots  represents the magnetic-field difference ratio $D$. The region of interest (ROI) is defined as a spherical volume with a diameter of $2.5~\rm{mm}$ centered at the vapor cell, and the insets show enlarged views of the central ROI.}
\label{fig:Bdis}
\end{figure}
}

\newcommand{\FigBsim}{
\begin{figure}[t]
\centering
\includegraphics[width=1\columnwidth]{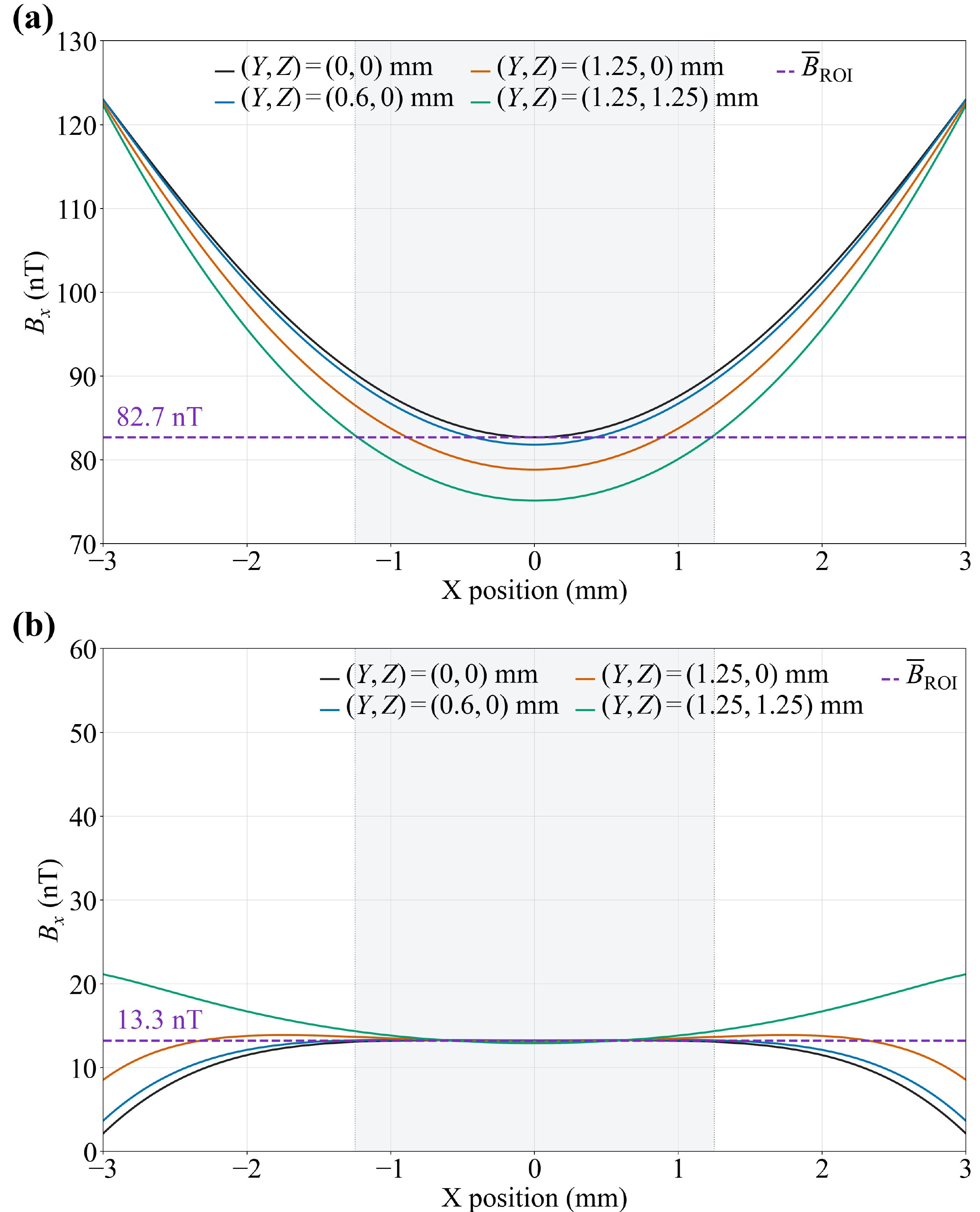}
\caption{Simulated $B_x$ distributions along the $X$ direction at different transverse positions $(Y,Z)$ for (a) SH and (b) BP configurations. Each curve represents a line cut at a fixed transverse position, corresponding to $(Y,Z)=(0,0)$, $(0.6,0)$, $(1.25,0)$, and $(1.25,1.25)~\rm{mm}$. The purple dashed line indicates the mean magnetic field within the ROI, $\overline{B}_{\rm{x,ROI}}$.}
\label{fig:Bsim}
\end{figure}
}

\newcommand{\Figcom}{
\begin{figure}[t]
\centering
\includegraphics[width=1\columnwidth]{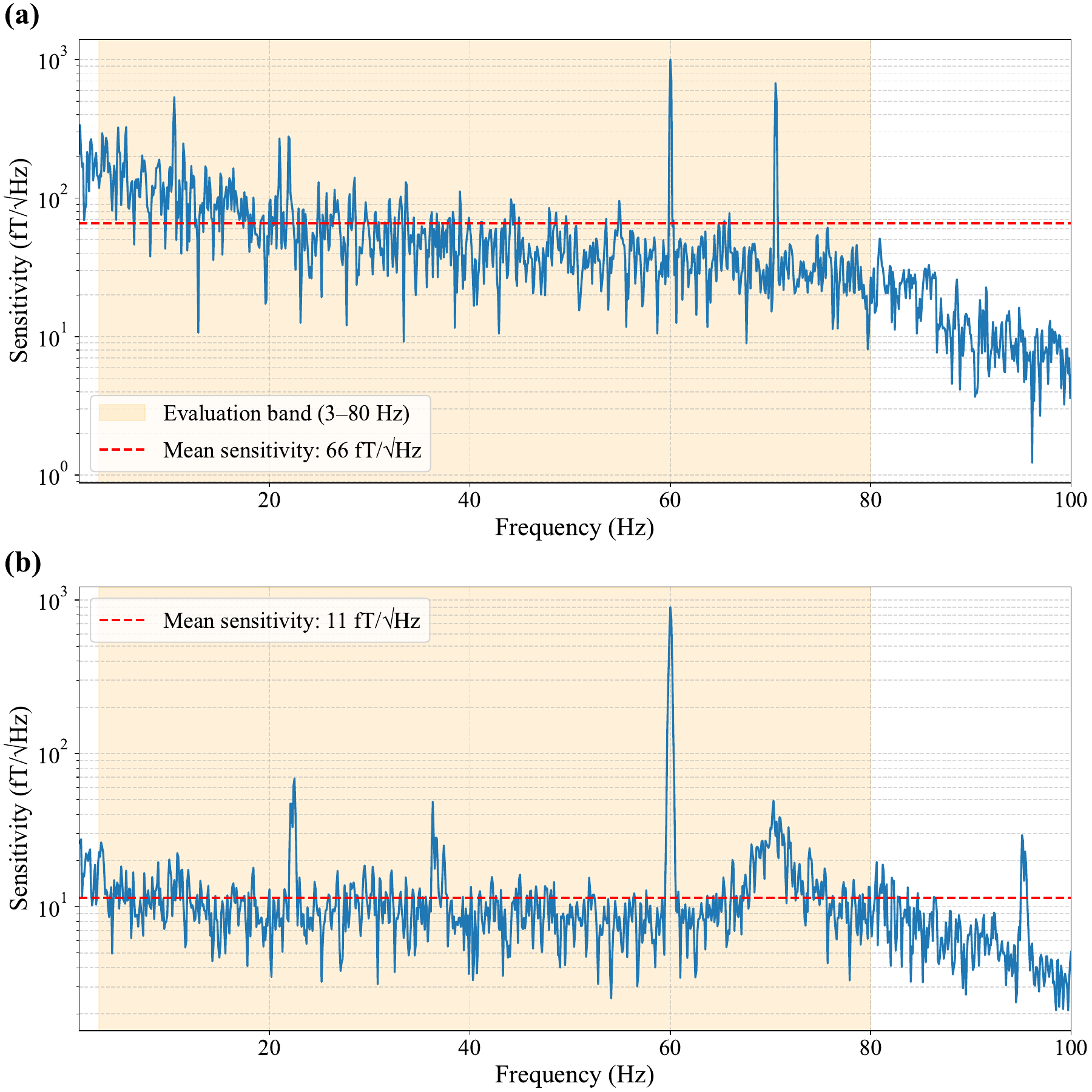}
\caption{Comparison of the magnetic-field sensitivity spectra between the miniaturized BP configuration in (a) and the benchtop system in (b). The shaded regions indicate the evaluation band from 3 to 80 Hz. The red dashed lines represent the mean sensitivity within this frequency range, corresponding to $66~\rm{fT}/\sqrt{\rm{Hz}}$ and $11~\rm{fT}/\sqrt{\rm{Hz}}$.}
\label{fig:com}
\end{figure}
}


\section{Introduction}
High-sensitivity magnetometry plays an important role in many scientific fields, including physics, biology, neuroscience, materials science, and geology. While low-temperature superconducting quantum interference devices (SQUIDs) have been used for the most demanding applications~\cite{SQUIDSreview}, recent advances in atomic magnetometers, now capable of subfemtotesla sensitivity~\cite{Kominis2003,Dang2010, Wasilewski2010}, have expanded the potential for ultrasensitive magnetic field detection. In contrast to SQUIDs, optically pumped magnetometers (OPMs) are all-optical atomic magnetometers based on the widely adopted pump-probe configuration~\cite{Kominis2003,Shah2010}. They offer key advantages such as thermal operation, high spatial resolution, and potential for miniaturization. Most OPMs use alkali atoms like potassium (K), rubidium (Rb), or cesium (Cs) due to their strong optical transitions and hyperfine structure suitable for optical pumping. Helium-4 ($^4$He) is also used due to its metastable triplet state, and its inert nature, high diffusion rates, and low spin relaxation rates make it well-suited for operation in high-temperature or high-field environments~\cite{Bertrand2021,Fourcault2021}. Recently, squeezed probe light with backaction evasion can enhance the sensitivity of spin-exchange-limited OPMs beyond the coherent-state limit~\cite{Troullinou2021,Troullinou2023}. Moreover, a deep neural network can extract full vector magnetic field information from optical rotation signals in a simplified scalar OPM setup, significantly reducing system complexity while maintaining high sensitivity~\cite{Meng2023}. These advances enable OPM-based magnetoencephalography (OPM-MEG), as well as other biomedical sensing and imaging applications, offering noninvasive, high-resolution magnetic field measurements with improved practicality~\cite{Borna2017,MEG2022,Aslam2023}. 

Atomic magnetometers are among the most sensitive magnetic field sensors, but their performance is often limited by spin-exchange collisions. However, this limitation can be overcome using the spin-exchange relaxation-free (SERF) technique~\cite{Allred2002,Savukov2005,Shah2009}. After years of development, SERF magnetometers have achieved an ultra-high sensitivity of sub-fT$/\sqrt{\rm{Hz}}$~\cite{Dang2010, Wasilewski2010}. To suppress spin-exchange relaxation and achieve optimal magnetic-field sensitivity, the spin-exchange rate must greatly exceed the Larmor precession frequency~\cite{Ledbetter2008,Zhang2019,Jiang2021}. This condition requires a high atomic number density and an extremely low magnetic field, realized through a heating system and a magnetic shielding system, respectively. The development of SERF magnetometers, including atomic sources, heating temperatures, and magnetic field sensitivity, is detailed in the review article~\cite{Li2018}. Since the sensitivity of OPMs is fundamentally linked to the volume of the atomic vapor cell, larger volumes generally enable higher signal-to-noise ratios (SNRs), miniaturization efforts must carefully balance size with performance~\cite{Tierney2019}.
Notable developments include the use of microfabricated vapor cells for compact designs~\cite{Shah2007}, atomic magnetometers utilizing a single elliptically polarized light achieving a sensitivity of 7~fT$/\sqrt{\rm{Hz}}$~\cite{Shah2009}, and single-beam configurations capable of high-sensitivity three-axis magnetic field measurements~\cite{tang2021}. Reference \cite{Limes2020} demonstrates a portable optical gradiometer using microfabricated alkali-vapor cells to detect biomagnetic signals from the human brain and heart in Earth's ambient magnetic field without shielding, achieving 16~fT$/\rm{cm}/\sqrt{\rm{Hz}}$ sensitivity.

To realize compact OPMs while maintaining high sensitivity, it is important to suppress atom-wall collisions. Buffer gas provides an effective means of slowing atomic diffusion to the cell walls, thereby reducing wall-induced relaxation. Here, we employ a high-buffer-pressure (500~Torr N$_2$) Rb vapor cell together with optical absorption detection.
This mode uses a single circularly polarized laser beam for both optical pumping and probing, eliminating the need for separate pump and probe paths and polarization-selective detection components. The pump field redistributes the atomic population among spin states through optical pumping, and this redistribution is detected as a change in the transmitted light intensity. Although the optical absorption mode generally offers a lower SNR than the optical rotation mode, it benefits from a significantly reduced system size and complexity. In this work, we achieved a sensitivity of 11~fT$/\sqrt{\rm{Hz}}$ using a small cell volume of only 0.027 cm$^3$ under optimized optical intensity and laser detuning conditions. 
We further developed a miniaturized OPM using a folded flexible printed circuit board (FPCB) to integrate the vapor cell, a compact heater, and magnetic-field coils into a cubic sensing head.
Two coil configurations, square-Helmholtz (SH) and biplanar (BP), were patterned directly on the FPCB and investigated to evaluate their magnetic-field uniformity and influence on sensor performance. The prototype was characterized using benchtop optical and readout components. Under optimized operating conditions, the miniaturized OPM achieved a sensitivity of 66~fT$/\sqrt{\rm{Hz}}$, demonstrating the potential of the FPCB-based architecture for compact, high-sensitivity OPMs.

\section{Theory and Methods}

An atom with a magnetic moment placed in an external magnetic field experiences a torque that causes its magnetic moment to precess around the field direction at the Larmor frequency, $\omega_L=\gamma B$, where $B$ is the magnetic field strength and 
$\gamma$ is the effective angular gyromagnetic ratio. The commonly quoted hyperfine-state gyromagnetic ratios $\gamma/2\pi$ for $^{87}\rm{Rb}$ and $^{85}\rm{Rb}$ are approximately $7~\rm{Hz/nT}$ and $4.7~\rm{Hz/nT}$, respectively. 
In the high-temperature SERF regime, rapid spin-exchange collisions average the spin precession between hyperfine manifolds, reducing the effective gyromagnetic ratio and suppressing spin-exchange relaxation, thereby narrowing the magnetic resonance linewidth~\cite{Zhang2019}.
In optical magnetometry experiments, the Larmor precession modulates the population distribution among Zeeman sublevels, affecting the optical response of the atomic ensemble. When a polarized resonant light interacts with atoms, optical pumping drives the population into dark states. The formation and stability of these dark states are highly sensitive to the Larmor frequency, and thus to the applied magnetic field.

\subsection{Magnetic Resonance Linewidth and Sensitivity}
\label{sec:2.1}
By scanning the magnetic field and monitoring changes in optical transmission, magnetic resonance features centered near zero field can be observed, as shown in Fig.~\ref{fig:zfr}. The full width of this magnetic resonance, $\Delta B$, is related to the effective decoherence time of the transverse spin polarization, $\tau$, via the gyromagnetic ratio,
\begin{equation}
\Delta B=\frac{2}{\gamma \tau}.
\label{eq:FWHM}
\end{equation}
OPM operation relies on several key physical processes, including optical pumping, Larmor precession, spin relaxation, and optical detection. A circularly polarized laser beam selectively excites atoms into a preferred spin orientation, generating a net spin polarization along the optical axis. In the presence of an external magnetic field $\vec{B}$, the collective spin polarization precesses about the field direction at the Larmor frequency.
Spin polarization relaxes due to a combination of mechanisms, including spin-exchange and spin-destruction collisions, magnetic-field inhomogeneity, wall collisions, and optical pumping itself. 
The spin coherence time is commonly characterized as $\tau=1/(R_p+1/T_2)$, where $R_p$ is the optical pumping rate, proportional to the light intensity, while $T_2$ denotes the transverse spin-relaxation time in the absence of light, including spin-destruction collisions with Rb and buffer gas N$_2$ atoms and diffusion relaxation~\cite{Shah2009}. 
Changes in spin orientation are read out optically via the absorption or transmission of the pump or a separated probe beam~\cite{Shah2009, Li2018}. In the zero-field resonance (ZFR) regime, the transmission signal is proportional to the $x$-component of the spin polarization, which is commonly described by~\cite{Tierney2019}, 
\begin{equation}
S_x(B)=\frac{S_0}{1+\left(\gamma \tau B\right)^2},
\label{eq:ZFR}
\end{equation}
where $S_0$ denotes the resonant signal amplitude and the full-width at half maximum (FWHM) is $\Delta B=2/\gamma \tau$ as shown in Eq.~(\ref{eq:FWHM}).
This Lorentzian lineshape reflects the coherent spin response of the atomic ensemble to a transverse magnetic field. As $B$ increases from zero, the spin polarization is gradually dephased by Larmor precession, reducing the transmission signal. 
The resonance linewidth $\Delta B$ is closely related to the spin coherence time $\tau$, with a narrower linewidth corresponding to longer-lived spin polarization and improved magnetometric sensitivity. Increasing the optical pumping rate, typically by raising the light intensity, increases the atomic spin polarization and hence the optical signal contrast. However, when the pumping rate becomes too large, light-induced relaxation and power broadening shorten $\tau$, leading to a broader magnetic resonance and degraded sensitivity. Consequently, an optimal pumping rate exists that balances enhanced spin polarization and signal contrast against linewidth broadening.

In the SERF regime, the rate of spin-exchange collisions between alkali atoms, $R_{\rm{SE}}$, greatly exceeds the Larmor precession frequency ($R_{\rm{SE}} \gg \omega_L$)~\cite{Allred2002,Savukov2005,Ledbetter2008,Shah2009}. Rapid spin-exchange interactions continuously mix the hyperfine states, effectively averaging over the different hyperfine manifolds. 
As a result, spin-exchange relaxation is strongly suppressed, reducing its contribution to the magnetic resonance linewidth, while rapid spin averaging enables extremely narrow zero-field resonances.

\FigZFR
The magnetic-field sensitivity of an OPM quantifies the smallest detectable magnetic-field variation within a given measurement bandwidth. In practice, a small oscillating magnetic field is applied to modulate the spin precession, and the transmission signal is demodulated using a lock-in amplifier (LIA). This modulation and demodulation scheme extracts the dispersive derivative of the magnetic resonance. The sensitivity is then determined by both the noise level of the detected signal and the slope of this dispersive signal, and is commonly expressed as
\begin{equation}
\delta B=\frac{\delta S}{\left|\frac{dS}{dB}\right|},
\label{eq:B}
\end{equation}
where $\delta S$ denotes the root-mean-square (RMS) noise of the detected signal and $\frac{dS}{dB}$ is the slope of the signal with respect to the magnetic field. 
The noise $\delta S$ is typically characterized in the frequency domain using a fast Fourier transform (FFT), allowing the sensitivity to be expressed in units of fT$/\sqrt{\rm{Hz}}$. 
The slope $\frac{dS}{dB}$ is usually evaluated near the linear region of the magnetic resonance, where the response to small magnetic-field variations is maximal. By combining the measured noise spectral density and the resonance slope, the sensitivity spectrum of the OPM can be obtained, providing a quantitative evaluation of sensor performance under different operating conditions.

\subsection{Zero-Field Resonance}
In the zero-field optical absorption configuration, the presence of a weak transverse magnetic field, typically on the order of a few tens of nT, reduces the probe-beam transmission relative to the zero-field resonance, as shown in Fig.~\ref{fig:zfr}. This behavior can be understood in terms of optical pumping and dark-state formation. At zero magnetic field, the atomic sublevels remain degenerate, allowing circularly polarized $\sigma^+$ light to optically pump atoms into dark states. 
A transverse magnetic field induces Larmor precession, mixing the spin states and disrupting dark-state formation, thereby reducing the transmitted probe intensity.
The quantization axis is defined by the magnetic-field direction along the $z$-axis, orthogonal to the light-propagation direction along the $x$-axis. In this basis, the incident circular polarization is decomposed into multiple components that couple different Zeeman sublevels. This disrupts effective optical pumping into dark states, resulting in enhanced absorption. Even pT- or fT-level magnetic fields can be detected by monitoring the transmission.
The enhanced absorption under a weak transverse magnetic field is primarily due to coherent dark state formation, which is favored by the D$_1$ transition. Unlike the D$_2$ transition, the D$_1$ transition lacks closed cycling transitions, allowing atoms to be optically pumped more efficiently into coherent superpositions of ground-state magnetic sublevels that are decoupled from the light field and form dark states.
\Figsetup

\subsection{Zero-Field Resonance in a Buffer Gas Cell}
\label{sec:ZFRresults}

In practical OPM systems, buffer gases at high pressures are commonly inserted into the vapor cell to narrow the magnetic resonance linewidth~\cite{Happer1973,Wang2022}. A buffer gas, typically at a pressure of a few hundred torr, reduces the diffusion rate of rubidium atoms out of the laser beam and suppresses wall collisions, allowing the atoms to interact with the light field for a longer duration. This extended interaction time improves the efficiency of optical pumping and stabilizes the formation of dark states. In addition, buffer gases such as N$_2$ act as quenching agents, providing non-radiative decay channels that reduce radiation trapping and velocity-selective effects under strong optical pumping. 
Furthermore, pressure broadening substantially increases the optical linewidth, rendering the detailed hyperfine structure less critical for optical pumping.

\section{Setup}
\label{sec:setup}
\subsection{Atomic System and Magnetic Environment}

The experimental setup consists of a magnetic shield, a heating system, a laser source, and a data acquisition and analysis system, as shown in Fig.~\ref{fig:setup}. The glass cell used in this work is a microfabricated vapor cell made by Twinleaf, with external dimensions of $5 \times 5 \times 3.4~\text{mm}^3$ and internal dimensions of $3 \times 3 \times 3~\text{mm}^3$. It contains pure $^{87}\text{Rb}$ and is filled with 500~Torr of $\rm{N_2}$ buffer gas. The cell is mounted using non-magnetic polyetheretherketone (PEEK) fixtures and housed within a four-layer magnetic shield (Twinleaf: MS-1LF) consisting of three $\mu$-metal layers and one ferrite layer. The shield is equipped with three pairs of internal compensation coils. During the measurements, the coils are used to apply both a scanned DC magnetic field and an AC modulation field. 
The residual magnetic field is actively minimized to below 400~pT driven by a digital degaussing unit (Twinleaf: DG-10), as verified by a commercial zero-field magnetometer (QuSpin: QZFM Gen-3). 

However, several factors can induce unwanted magnetic fields, such as heating elements and temperature sensors. The cell temperature is maintained at 437~K, corresponding to the atomic density of 2.0$\times 10^{14}$~cm$^{-3}$, using a heater printed on a 150~$\rm{\mu}$m-thick hydroxide-filled ceramic PCB as shown in Fig.~\ref{fig:setup}(c). 
A non-magnetic silver-coated RTD (PT1000) is attached to the side of the vapor cell using UV-curable adhesive and serves as the feedback sensor for a proportional–integral (PI) temperature controller implemented on the FPGA, which regulates the heater drive power. The temperature stability is $\pm$0.05$^\circ$C. The heater features a double-sided and counter-running circuit design to cancel out the magnetic field. It is driven by a custom-built circuit based on an FPGA and an H-bridge circuit, which generates a high-frequency AC current at 351~kHz more than three orders of magnitude above the detection band (DC–100~Hz). This frequency was chosen to balance thermal efficiency and magnetic-noise suppression, minimizing current-induced magnetic interference in both temporal and spatial domains~\cite{Shah2009,Colombo2016FourChannel,Lu2020Heater,Liu2021ITO} . 
To reduce the residual magnetic field generated by the operating heater ($\sim$6~nT at the heater surface), a 1.5~mm-thick Al$_2$O$_3$ substrate is placed between the heater and the vapor cell. This configuration suppresses the field across the optical beam to approximately 2~nT.

\subsection{Laser Power Stabilization and Magnetic Field Modulation}
\label{sec:differential}

A probe beam is generated by a distributed Bragg reflector laser (DBR, Thorlabs: DBR795PN) tuned to the D$_1$ transition of $^{87}$Rb (795~nm ) and coupled into a polarization-maintaining (PM) fiber. It has a diameter of 2.6~mm, which increases the interaction volume and the fraction of polarized $^{87}$Rb atoms in the vapor cell. To stabilize the optical power incident on the cell, a feedback loop consisting of a homemade liquid-crystal variable retarder (LCVR), operated as a variable half-wave plate ($\lambda/2$), and a polarizing beamsplitter (PBS) is employed.
In addition, a fixed $\lambda/2$ is placed before the second PBS to adjust the polarization such that the incident beam is split into equal transmitted and reflected components. The reflected beam is monitored by a photodetector (PD; Thorlabs: PDA36A2), providing a calibrated measure of the optical power incident on the cell. The PD signal is compared with a predefined setpoint to generate an error signal, which is fed back to the LCVR. By adjusting the polarization rotation angle, the LCVR stabilizes the optical power transmitted through the PBS toward the vapor cell.
A quarter-wave plate ($\lambda/4$) is used to convert the probe beam into circular polarization before it enters the vapor cell. A reference cell filled with natural Rb atoms is applied as the frequency reference. During the experiments, the probe laser frequency was scanned across Rb D$_1$ transition over a total range of 17~GHz. For OPM magnetic spectral measurements, the optical intensity incident on the vapor cell was varied between 1.5 $\sim$ 2.5~mW/cm$^2$.

Before acquiring the magnetic-resonance curves, the residual field was minimized after degaussing, and the Helmholtz coils were used to generate the scanned DC and modulated AC fields.
The AC field is modulated at 997~Hz driven by the lock-in amplifier (LIA; Zurich Instruments: MFLI).
The DC magnetic field was scanned at 0.1~Hz using the same LIA, with a calibrated field conversion factor of 2.9~nT/mV, as determined using the magnetometer QZFM Gen-3.
The transmitted probe signal is detected by a PD and sent to the LIA. The signal is demodulated at the modulation frequency and subsequently passed through a low-pass filter with a cutoff frequency of 100~Hz for continuous data acquisition. The measured zero-field resonance and dispersion curve are shown in Fig.~\ref{fig:zfr}.

\section{Results and Discussion} \label{sec: results}
\Figspectrum
\subsection{Transmission Spectra in a Buffer Gas Cell}
The typical transmission spectra are shown in Fig.~\ref{fig:spectrum}. With $^{85}$Rb and $^{87}$Rb isotopes, the transmission spectrum of Ref.~Cell exhibits clearly resolved dips corresponding to the hyperfine ground states. As shown by the black line, the two deeper central dips are attributed to $^{85}$Rb, whereas the shallower dips on the two spectral wings correspond to the $^{87}$Rb hyperfine transitions from the $F=2$ and $F=1$ ground states. The hyperfine splitting of the excited states is larger than the Doppler linewidth, so these transition lines remain well resolved. We define zero detuning with respect to the $F=2\rightarrow F'=1$ transition. This spectrum serves as a frequency reference for calibrating the probe laser detuning throughout the measurements.

For pure $^{87}$Rb mixed with a 500 Torr N$_2$ buffer gas, the strong pressure broadening merges the absorption lines into a single, substantially broadened profile~\cite{SHI2019,McWilliam2024}. The buffer gas also induces a pressure shift, typically leading to a redshift of several GHz. This broadened and smoother absorption spectrum can improve the stability of optical magnetometry. In addition, the involved hyperfine transitions, which have different transition strengths, contribute differently to the spectrum and give rise to the characteristic dual-feature transmission profile.

Furthermore, the transmission spectrum is intensity dependent. As the optical intensity increases, e.g., the red curve at 11~mW/cm$^2$, the $F=2$ ground state near $-3~$GHz shows deeper absorption compared to the $F=1$ ground state near 4~GHz. As the optical intensity decreases, the relative contributions of power broadening and optical pumping efficiency differ for the two transitions, resulting in different resonance responses.
The spectral depth reflects the optical pumping rate, and increasing it enhances both atomic spin polarization and the relaxation rate.
Understanding this dual-feature and intensity-dependent evolution is important for optimizing zero-field OPM operation. 
Theoretically, the distinct saturation intensities and power-broadening behaviors of the $F=1$ and $F=2$ hyperfine transitions imply that optical pumping on each ground state can produce unequal response, since the optical pumping rate, spin-polarization formation, and relaxation processes depend on the transition and optical intensity.

\subsection{Effects of Laser Intensity and Detuning on Spectral Features}
\Figdensity
The magnetic field sensitivity is defined as $\delta B=\frac{\delta S}{|\frac{dS}{dB}|}$. The pumping rate must be optimized to balance improved spin polarization and signal contrast with the accompanying linewidth broadening. We first focus on the magnetic-resonance response to examine how optical intensity and laser detuning affect its key characteristics, including the linewidth and peak height of the zero-field resonance and the slope of the differential signal.
Through a systematic study of the effects of laser intensity and detuning, we present the characteristics of our OPM in Fig.~\ref{fig:density}. 
As shown in Fig.~\ref{fig:density}(a), the extracted linewidths (FWHM) are relatively narrow at detunings of approximately -3.9 and 2.8~GHz. 
We further extract measurements at two representative intensities (1.5 and 1.9 mW/cm$^2$) and present them in Fig.~\ref{fig:density}(b). 
Both intensities show a dual-structure response.
In addition, Fig.~\ref{fig:density}(c) shows that the linewidths increase with the intensity. As discussed in Sec.~\ref{sec:2.1}, $\Delta B=\frac{2}{\gamma \tau}$ and $\tau=1/(R_p+1/T_2)$. 
Under our experimental condition, the optical pumping rate $R_p\gg1/T_2$, the reciprocal of the transverse spin-relaxation time. 
At sufficiently high optical intensity ($>1.5$~mW/cm$^2$), where the probe beam is no longer fully absorbed, the magnetic resonance linewidth becomes approximately proportional to the optical intensity, as indicated by the brown dashed guide line.
Further increasing the optical intensity induces power broadening, leading to a steeper linear increase in the linewidth, as indicated by the black dash-dotted line.

Similarly, the zero-field resonance peak heights and differential slopes are extracted as functions of the laser intensity and detuning, as shown in Figs.~\ref{fig:density}(d)--\ref{fig:density}(i).
The detuning dependence of the peak height is presented in Fig.~\ref{fig:density}(e). At both intensities, the peak height exhibits a clear dual-structure response near the two resonance regions. As the optical intensity increases, the overall peak height becomes larger, while the dual-structure feature remains visible.
The intensity dependence of the peak height in Fig.~\ref{fig:density}(f) can also be divided into two regimes. At low optical intensities, the probe beam is strongly absorbed during propagation through the atomic vapor, allowing only a small fraction of the beam to be transmitted through the atomic ensemble. This behavior was directly observed via the fluorescence signal captured by a CCD camera. Similar effects have also been reported in previous studies using pump-probe configurations, in which the distribution of spin polarization along the laser beam direction was measured~\cite{Colombo2016FourChannel,Zhao2019}. The reduction of laser intensity is given by 
\begin{equation}
\frac{dI(x)}{dx}=-n\sigma I(x)(1-P(x)),
\label{eq:reduction}
\end{equation}
where $n$ is the atomic density, $\sigma$ is the absorption cross section, and $P(x)$ is the spin polarization as function of atomic position $x$.
As a result, the detected signal increases slowly with the incident intensity, leading to a relatively weak increase in the peak height.
Above a certain threshold intensity ($2.0$~mW/cm$^2$), the absorption becomes saturated and the probe beam is no longer completely absorbed.
With increasing optical intensity, optical pumping becomes more significant, leading to stronger atomic polarization and a correspondingly steeper increase in the peak height.

The differential slope is extracted via a lock-in amplifier with field modulation. The detuning dependence of the slope exhibits a similar dual-structure response, indicating that all three quantities are governed by the same underlying resonance features. Here we focus on how the optical intensity varies the slope as shown in Fig.~\ref{fig:density}(i). 
Since the slope is related to the ratio between the peak height and the linewidth, its intensity dependence follows the same two-regime behavior discussed above. In the present measurement, the slope generally increases with optical intensity, which is favorable because the slope appears in the denominator of the magnetic-field sensitivity. However, the sensitivity is also determined by the noise level in the numerator. Therefore, although increasing the optical intensity improves the differential slope, the overall sensitivity also depends on the noise level, whose intensity dependence must be examined next.

\subsection{Sensitivity}
\Figsensitivity
We now turn to the noise term, $\delta S$, which represents the amplitude of fluctuations in the detected signal and plays an important role in determining the sensitivity.
Low intensity results in insufficient optical pumping and weak signals, whereas high optical intensity causes power broadening and increased noise.
Following the analysis in Fig.~\ref{fig:density}, we further evaluate the sensitivity by including the measured noise. 
The noise density and the corresponding sensitivity are shown as functions of optical intensity and laser detuning in Fig.~\ref{fig:sensitivity}.

For noise measurement, the field scan was then stopped, and the demodulated output from the LIA was recorded. The amplitude spectral density (ASD) of the time-domain signal was calculated using Welch's method with a Hann window and 50\% overlap. The noise density $\delta S$ was obtained by averaging the ASD over the 3--80~Hz frequency range. 
The OPM sensitivity was then calculated as $\delta B=\frac{\delta S}{|\frac{dS}{dB}|}$.

At low optical intensities, the measured noise ASD remains nearly insensitive to both laser detuning and optical intensity, as shown in Figs.~\ref{fig:sensitivity}(b) and \ref{fig:sensitivity}(c). 
However, with increasing optical intensity, a dual-structure feature emerges in the detuning-dependent noise spectrum, accompanied by a dramatic rise in noise at higher intensities. 
In particular, when the optical intensity increases beyond the knee point near 2.0~mW/cm$^2$, the noise increases rapidly. This behavior indicates that additional intensity-dependent noise mechanisms become significant in the high-intensity regime.

The sharp increase in noise observed at high probe intensity is closely associated with the enhanced transmission peak height, as shown in Fig.~\ref{fig:density}(f).
When the probe intensity is increased from the knee point to 2.4~mW/cm$^2$, the peak height increases by approximately a factor of 3.4, amplifying the noise amplitude. However, the measured noise increases by a factor of about 20 over the same intensity range in Fig.~\ref{fig:sensitivity}(c). 
The remaining noise increase is likely associated with the circularly polarized single-beam pump-probe mechanism. In this configuration, increasing the probe intensity modifies the optical pumping rate, spin polarization, and lock-in response, thereby enabling probe-intensity fluctuations to couple into spin-polarization fluctuations and demodulated-signal noise. This mechanism may account for the observed noise increase beyond that expected from the peak transmission enhancement alone.

By combining the measured noise with the differential slope, we obtain the OPM sensitivity, as shown in Figs.~\ref{fig:sensitivity}(d)--(f). The best sensitivity is obtained near the $F=1$ and $F=2$ resonance regions at an optical intensity of 1.9~mW/cm$^2$. The intensity dependence of the optimized sensitivity is summarized in Fig.~\ref{fig:sensitivity}(f). In the weak-intensity regime, the noise remains nearly insensitive to optical intensity, while the differential slope increases with intensity, leading to an improvement in sensitivity. However, further increasing the intensity introduces additional noise, which degrades the sensitivity. Therefore, the optimal intensity reflects a trade-off between differential slope and noise.
The best sensitivity of 11~fT$/\sqrt{\rm{Hz}}$ is achieved near the $F=2$ resonance transition.

\section{Miniaturized Sensor Prototype}

The previous section described the benchtop OPM system and systematically characterized how the field sensitivity depends on laser detuning and optical intensity. Based on these results, we developed a miniaturized sensor prototype to evaluate the feasibility of integrating the single-beam absorption-mode OPM into a compact sensor head. Here, we describe the miniaturized sensor structure and performance relative to the benchtop system.

\FigFPCB
\subsection{Integrated Sensor Architecture and Experimental Configuration}

The design and assembly of the miniaturized sensor prototype are shown in Fig.~\ref{fig:FPCB}. The custom FPCB is divided into three main functional regions: main sensing (labeled I), heating (II), and laser integration (III). 
The main sensing region (I) houses the vapor cell and magnetic-field coils. It also provides electrical interfaces for the temperature sensor and the PDs; the PDs are reserved for future integration and were not used in this work.
The heating region (II) provides thermal control for the vapor cell, whereas the reserved laser-integration region (III) contains the electrical interface for the laser diode. Figure~\ref{fig:FPCB}(b) shows the partially assembled sensing head. 
Six support plates made of FR-5 glass-fiber-reinforced epoxy laminate have been attached to the backside of the FPCB to provide mechanical support and define the geometry of the folded sensing head. 
A compact square heater with a side length of 5.5~mm, compared with 6.5~mm for the heater used in the benchtop system, was installed in the cell holder.
A UV-curable adhesive was applied to secure the microfabricated vapor cell and cell holder to the surrounding FPCB. 
After assembly, the sensing region was folded into a compact sensing head, as shown in Fig.~\ref{fig:FPCB}(c), with an outer dimension of  $12~\rm{mm}$ on each side.

For the characterization of the miniaturized sensor prototype, the optical excitation and detection were performed using the same DBR laser, polarization optics, PD, and electronics for lock-in amplification and data acquisition as those used in the benchtop setup. 
The miniaturized Helmholtz coils were designed and patterned directly on the FPCB. 
Magnetic-field gradients within the optical interaction volume can significantly degrade sensor performance. This effect can be partially suppressed by reducing the laser-beam diameter, which confines the interaction region to the more homogeneous central region of the magnetic field.
However, according to the sensitivity scaling of OPMs, the smallest measurable magnetic field can be expressed as~\cite{Tierney2019}
\begin{equation}
B_{\min}=\frac{1}{\gamma\sqrt{n\tau Vt}},
\end{equation}
where $V$ is the optical interaction volume, and $t$ is the measurement time. The sensitivity limit is strongly affected by the number of atoms participating in the measurement, which is proportional to $nV$. Reducing the beam size decreases the number of atoms contributing to the light–atom interaction and consequently raises the fundamental sensitivity limit. 
The interaction volume can therefore be increased by designing a larger region with a homogeneous magnetic field.

\subsection{Magnetic Coils Designed for Integrated Sensor}

To evaluate the effect of the integrated coil geometry on the performance, two coil configurations were were simulated, and experimentally investigated. 
The first configuration, referred to as the square-Helmholtz (SH) configuration, used rounded-square Helmholtz coils patterned on the FPCB. The second configuration, referred to as the biplanar (BP) configuration, used integrated biplanar coils.
In both configurations, the integrated coils provided all magnetic fields required for residual-field compensation, DC magnetic-field scanning, and AC modulation. 
The magnetic-field distributions are simulated to evaluate the influence of the coil geometry on the homogeneous-field region, as shown in Fig.~\ref{fig:Bdis}. The coil current is set to $1~\rm{mA}$.  
The region of interest (ROI) is defined as a spherical volume with a diameter of $2.5~\rm{mm}$ centered at the vapor cell. The magnetic-field difference ratio, $D$, is defined as
\begin{equation}
D=\frac{|\bf{B}|-\overline{|\bf{B}|}_{\rm{ROI}}}
{\overline{|\bf{B}|}_{\rm{ROI}}}\times100\%,
\end{equation}
where $|\bf{B}|$ is the local magnetic-field magnitude and $\overline{|\bf{B}|}_{\rm{ROI}}$ is the average magnetic-field magnitude within the ROI.
\FigBdis
\FigBsim

Figure~\ref{fig:Bdis} shows the simulated spatial magnetic-field distributions with the color representing the difference ratio $D$. The insets in the upper-left corners show enlarged views of the central ROI. The BP configuration exhibits smaller spatial variations within the ROI, indicating improved magnetic-field homogeneity around the center of the vapor cell.
Figure~\ref{fig:Bsim} further shows the simulated $B_x$ distributions along the $X$ direction at different transverse positions $(Y,Z)$. In the SH configuration, the $B_x$ profiles exhibit noticeable curvature, indicating substantial spatial variation in the magnetic field. The mean magnetic field within the ROI, $\overline{B}_{\rm{x,ROI}}$, is $82.7~\rm{nT}$, with an estimated magnetic-field variation of $13.8\%$. In contrast, the BP configuration exhibits substantially flatter $B_x$ profiles and yields $\overline{B}_{\rm{x,ROI}}\approx13.3~\rm{nT}$, with a field variation of only $2.6\%$. These results demonstrate that the BP configuration provides a considerably more homogeneous magnetic-field distribution.

\subsection{Performance and Comparison with the Benchtop System}

The performance of the two miniaturized coil configurations was subsequently evaluated experimentally. For the SH configuration, the limited homogeneous-field region required the laser-beam diameter to be reduced to $1.6~\rm{mm}$. After optimization, the best magnetic-field sensitivity obtained with the SH configuration was $1.2~\rm{pT}/\sqrt{\rm{Hz}}$.
For the BP design, the laser beam can be increased to $2.6~\rm{mm}$, identical to that used in the benchtop system. 
Under the optimized operating conditions, the BP configuration achieved a sensitivity of $66~\rm{fT}/\sqrt{\rm{Hz}}$.
Even with the reduced beam size, the magnetic-field uniformity within the illuminated volume of the SH configuration remains lower than that provided by the BP coils. The resulting field gradients introduce additional inhomogeneous broadening and degrade the resonance slope. These results demonstrate that increasing the homogeneous-field volume is particularly important for maintaining high sensitivity in a miniaturized vapor-cell OPM.

Having identified the BP configuration as the preferred design, we compare its sensitivity spectrum with that of the benchtop system in Fig.~\ref{fig:com}.
The sensitivity was evaluated over the frequency range from 3 to 80 Hz, as indicated by the shaded region. Prominent narrowband interference peaks within this range were excluded from the averaging procedure to avoid overestimating the broadband sensitivity level. The sensitivity was $66~\rm{fT}/\sqrt{\rm{Hz}}$, corresponding to a sixfold degradation in sensitivity after miniaturization.
The BP configuration exhibits a generally higher magnetic-field-equivalent noise level over most of the evaluation band. The noise level was approximately four times higher than that of the benchtop system, with a particular increase at frequencies below $20~\rm{Hz}$. This low-frequency noise increase is expected to be mainly associated with the modified heating structure in the miniaturized sensor. Both spectra also contain several narrow spectral peaks, including a prominent component near $60~\rm{Hz}$, which are mainly associated with narrowband environmental or electrical interference.
The remaining degradation in sensitivity is mainly attributed to magnetic-resonance broadening. The resonance linewidth of the BP configuration was 50\% broader than that of the benchtop system, consistent with the residual field gradients of the BP coils, which are larger than those of the shield coils used in the benchtop system. A slight distortion of the resonance line shape was also observed, providing further evidence of spatial magnetic-field nonuniformity. 
\Figcom

\subsection{Outlook}
Future development will focus on further integrating the optical and detection components into the FPCB-based sensor platform, including the direct incorporation of both the PD and light source into the sensor head.
Since the vapor cell operates at temperatures up to $160^\circ\rm{C}$ and the PD will be located only a few millimeters away, the elevated local temperature may increase detector thermal noise and degrade the signal-to-noise ratio.
In addition, the light source will be replaced by an unpackaged vertical-cavity surface-emitting laser (VCSEL) for further miniaturization. Although the absorption spectrum is substantially broadened by the high buffer-gas pressure and high vapor-cell temperature, the influence of the broad laser linewidth of $120~\rm{MHz}$ should still be taken into account when evaluating the magnetometer performance. Moreover, the miniaturized VCSEL system may lead to larger optical-intensity fluctuations and consequently increase the optical noise of the magnetometer.

Further optimization of the integrated magnetic-field coils is also possible. According to the simulation results, replacing the single-current configuration with a two-current design and appropriately adjusting the current amplitudes and directions can further improve the magnetic-field uniformity. With the optimized current distribution, the magnetic-field difference ratio $D$ within the ROI can be reduced to below $1\%$, providing a potential route toward further improvement of the sensitivity of the fully integrated sensor.

\section{Conclusions}
We have developed a compact OPM system consisting of a laser source, a microfabricated vapor cell containing pure $^{87}$Rb with 500 Torr of N$_2$ buffer gas, a four-layer magnetic shield, and Helmholtz coils for controlling magnetic field. The system operates in absorption mode with a single circularly polarized beam. Magnetic field sensitivity is derived by analyzing the signal slope and noise spectrum. We systematically investigate the dependence of sensitivity on laser detuning and optical intensity, revealing a trade-off between peak height enhancement and noise increase. The best sensitivity of 11~fT$/\sqrt{\rm{Hz}}$ is achieved with the cell volume of 0.027~cm$^3$.
Furthermore, we developed a miniaturized OPM prototype using a folded FPCB to integrate the vapor cell, heater, and magnetic-field coils into a cubic sensing head of 12 mm on each side. Among the two coil configurations investigated, the BP design provided improved magnetic-field homogeneity and achieved a sensitivity of 66~fT$/\sqrt{\rm{Hz}}$. The remaining performance gap relative to the benchtop system highlights the importance of further reducing magnetic-field gradients within the optical interaction region. Future work will focus on integrating the light source and photodetector into the FPCB-based platform toward a fully integrated sensor.
This compact setup enables low-noise magnetic-field detection and provides a promising platform for miniaturized, multi-axis OPMs suited to biomedical and portable applications. 
The use of thermal atomic ensembles further supports its integration into chip-scale sensor platforms.



\bibliographystyle{IEEEtran}
\bibliography{OPMreference.bib}

\end{document}